\documentclass[useAMS,usenatbib,usegraphicx]{mn2e}

\title[Reddenings of supergiants and Cepheids]
{Reddenings of FGK supergiants and classical Cepheids from spectroscopic data}
\author[Kovtyukh, Soubiran, Luck, Turner, Belik, Andrievsky \& Chekhonadskikh]
{V.~V.~Kovtyukh$^1$\thanks{Email: val@deneb1.odessa.ua}, C.~Soubiran$^2$,
  R.~E.~Luck$^3$, D.~G.~Turner$^4$, S.~I.~Belik$^1$,
 \newauthor S.~M.~Andrievsky$^{1,5}$ and F.~A.~Chekhonadskikh$^{5}$ \\ \\
  $^1$Astronomical Observatory, Odessa National University, T.~G. Shevchenko Park, 65014, Odessa, 
      Ukraine \\
  $^2$Universit\'e de Bordeaux - CNRS - Laboratoire d'Astrophysique de Bordeaux, BP 89, 33270
    Floirac, France \\
  $^3$Department of Astronomy, Case Western Reserve University, 10900 Euclid Avenue, Cleveland, 
      OH 44106-7215, U.S.A. \\
  $^4$Department of Astronomy and Physics, Saint Mary's University, Halifax, Nova Scotia, B3H 3C3, 
      Canada \\
  $^5$Department of Astronomy, Odessa National University, T.~G.~Shevchenko Park, 65014, Odessa, 
      Ukraine \\
}
\date{Accepted 2008 June 25. Received 2008 June 25; in original form 2008 May 30}
\pagerange{\pageref{firstpage}--\pageref{lastpage}} \pubyear{2008}

\begin{document}

\label{firstpage}

\maketitle

\begin{abstract}
Accurate and homogeneous atmospheric parameters ($T_{\rm eff}$, $\log g$, $V_{\rm t}$, [Fe/H]) are derived for 74 FGK non-variable supergiants from high-resolution, high signal-to-noise ratio, echelle spectra. Extremely high precision for the inferred effective temperatures (10--40 K) is achieved by using the line-depth ratio method. The new data are combined with atmospheric values for 164 classical Cepheids, observed at 675 different pulsation phases, taken from our previously published studies. The derived values are correlated with unreddened {\it B--V} colours compiled from the literature for the investigated stars in order to obtain an empirical relationship of the form: $(B-V)_{0} = 57.984 - 10.3587(\log T_{\rm eff})^{2} + 1.67572(\log T_{\rm eff})^{3} - 3.356\log g + 0.0321V_{\rm t} + 0.2615[{\rm Fe/H}] + 0.8833(\log g)(\log T_{\rm eff})$. The expression is used to estimate colour excesses E({\it B--V}) for individual supergiants and classical Cepheids, with a precision of $\pm0.05$ mag. for supergiants and Cepheids with $n=1-2$ spectra, reaching $\pm0.025$ mag. for Cepheids with $n>2$ spectra, matching uncertainties for the most sophisticated photometric techniques. The reddening scale is also a close match to the system of space reddenings for Cepheids. The application range is for spectral types F0--K0 and luminosity classes I and II.
\end{abstract}


\section{Introduction}
The Cepheid period-luminosity (PL) relation continues to play a key role in the determination of distances within the Local Group and to nearby galaxies. The absolute calibration of the PL relation requires reliable distance measurements for calibrating Cepheids as well as accurate corrections for the effects of interstellar reddening and extinction. Galactic Cepheids are often heavily obscured, the average colour excess E({\it B--V}) being of order 0.5 mag. Here we propose a new method of accurate colour excess determination that relies on spectroscopically determined stellar parameters. 

The traditional methods of establishing the interstellar reddening of Cepheids have involved field reddenings derived from photometric and spectroscopic studies of stars surrounding the Cepheid, the use of reddening-independent spectroscopic indices, and period-colour relations calibrated from spectroscopic analyses of a number of well-studied Cepheids or supergiants. Field reddenings are fairly
reliable provided that the surrounding stars are unaffected by circumstellar reddening, although the number of suitable objects has tended to be restricted to cluster Cepheids and visual binaries \citep[see][]{tu95,tu01,lc07}. Reddening-independent indices include the $\Gamma$ index \citep{kr60,sj89}, which measures the depression in stellar spectra caused by the {\it G} band of CH ($\lambda 4305$), and the {\it KHG} index \citep{mp69,mc70,fe72,tu87}, which combines narrow band photometry of Ca II K ($\lambda 3933$), Balmer H$\delta$ ($\lambda 4101$), and the {\it G} band. Narrow band photometry of Balmer H$\beta$ ($\lambda 4861$) has also been used as a temperature indicator for FGK stars, but has been susceptible to problems arising from colour dependences introduced by mismatches in the effective
wavelengths of the narrow and wide H$\beta$ filters employed \citep{st79}.

Most Cepheid reddenings have been derived from period-colour relations that are tied to specific Cepheid or supergiant calibrators, in which specific continuum or spectral line features (line ratios, for example) are linked to stellar atmosphere models to infer effective temperatures for the stars \citep[e.g.,][]{fe87,sl90}. The direct link between effective temperature and broad band ({\it B--V})$_0$ colour has been demonstrated very effectively by \citet{gr92} for older, published, stellar atmosphere models, and has been used successfully to test the open cluster calibration of the PL relation \citep{tb02}. As argued by \citet{fc99}, {\it B--V} colours for Cepheids appear to be more closely indicative of stellar temperatures than indices that include a near-infrared magnitude. But recent changes to stellar atmosphere models \citep{ku92} may affect the effective temperature scale for FGK supergiants significantly enough that a recalibration is necessary. That is the purpose of the present study.

Cepheids also undergo significant changes in effective temperature during their pulsation cycles, which makes it essential to track such changes accurately through close monitoring of individual variables.
Photometric monitoring of Cepheids has resulted in a large database of published results for many different colour systems \citep{be07}, but the availability of high resolution spectra for such stars throughout their cycles has been extremely limited, until now. A goal of the present study is therefore to exploit a recently developed collection of high resolution spectra for Galactic Cepheids in order to present accurate intrinsic parameters for all of the studied objects.

\section{Observations}
The spectra of the FGK supergiants studied here were obtained using the 1.93\,m telescope of the Haute-Provence Observatoire (France) equipped with the echelle spectrograph ELODIE \citep{ba96} and retrieved from its on-line archive of spectra \citep{mou04}. The resolving power for the observations was $R=42\,000$ over the wavelength interval 4400--6800 \AA, with a signal-to-noise ratio for each spectrum of S/N$\geq$100 (at 5500 \AA). Initial processing of the spectra (image extraction, cosmic ray removal, flatfielding, etc.) was carried out as described by \citet{ka98}.

We also made use of spectra obtained with the Ultraviolet-Visual Echelle Spectrograph (UVES) instrument at the Very Large Telescope (VLT) Unit 2 Kueyen \citep{ba03}. All supergiants were observed in two instrumental modes, $Dichroic 1$ (DIC1) and $Dichroic 2$ (DIC2), in order to provide almost complete coverage of the wavelength interval $3000-10\,000$ \AA. The spectral resolution is about 80\,000, and for most of the spectra the typical S/N ratio is 300--500 in the {\it V} band.

For classical Cepheids in our sample we have used our previously published results \citep{ka99,an02a,an02b,an02c,lu03,an05,ko05,lu06}. Multiphase observations of program Cepheids were carried out using the Struve 2.1-m reflector and Sandiford echelle spectrograph at McDonald Observatory.
The nominal resolving power was 60\,000 with a total spectral range covering about $1000-1200$ \AA. The primary wavelength region was centered at 6200 \AA. Exposure times were selected to ensure an optimal S/N ratio of about 100. 
 
Additional processing of the spectra (continuum level location, measuring of line depths and equivalent widths) was carried out using the DECH20 software \citep{ga92}. Line depths R$_{\lambda}$ were measured by means of Gaussian fitting.

\section{Atmospheric parameters for FGK supergiants and classical Cepheids}
Effective temperatures for our program stars were established from the processed spectra using the method developed by \citet{ko07} that is based upon $T_{\rm eff}$--line depth relations. The technique can establish $T_{\rm eff}$ with exceptional precision. It relies upon the ratio of the central depths of two lines that have very different functional dependences on $T_{\rm eff}$, and uses tens of pairs of lines for each spectrum. The method is independent of interstellar reddening, and only marginally dependent on the individual characteristics of stars, such as rotation, microturbulence, metallicity, etc.

Briefly, a set of 131 line ratio--$T_{\rm eff}$ relations was employed, with the mean random error in a single calibration being 60--90\,K (40--45\,K in most cases and 90--95\,K in the least accurate cases). The use of $\sim$70--100 calibrations per spectrum reduces the uncertainty in $T_{\rm eff}$ to 10--20\,K for spectra with S/N ratios greater than 100, and 30--50\,K for spectra with S/N ratios less than 100. Although the internal error for each $T_{\rm eff}$ determination appears to be small, a systematic shift of the zero-point in the $T_{\rm eff}$ scale may exist. Nevertheless, an uncertainty in zero-point (if it exists) can affect the absolute abundances derived for each program star. It is relatively unimportant for abundance comparisons between stars in the sample. For the majority of supergiants and Cepheids we obtain error estimates for $T_{\rm eff}$ that are smaller than 10--40 K.

The microturbulent velocities, $V_{\rm t}$, and surface gravities, $\log g$, were derived using a modification of the standard analysis proposed by \citet{ka99}. As described there, the microturbulence is determined from the Fe~II lines rather than the Fe~I lines, as in classical abundance analyses. The surface gravity is established by forcing equality between the total iron abundance obtained from both Fe~I and Fe~II lines. Typically with this method the iron abundance determined from Fe~I lines shows a strong dependence on equivalent width (NLTE effects), so we take as the proper iron abundance the extrapolated total iron abundance at zero equivalent width.

Kurucz's WIDTH9 code was used with an atmospheric model for each star interpolated from a grid of models  calculated with a microturbulent velocity of 4 km s$^{-1}$. At some phases Cepheids can have microturbulent velocities deviating significantly from that value; however, our previous test calculations suggest that changes in the model microturbulence over a range of several km s$^{-1}$ has an insignificant impact on the resulting element abundances. Typical results obtained for the Cepheid $\delta$ Cep over its cycle are shown for illustrative purposes in Fig. 1.

Final results for our determinations of $T_{\rm eff}$, $\log g$, $V_{\rm t}$ (in km s$^{-1}$), and [Fe/H] for FGK supergiants in our sample are given in Table 1. Typical uncertainties in the cited values are $\pm(10-40)$ K in $T_{\rm eff}$, $\pm0.1$ in $\log g$, and $\pm0.5$ km s$^{-1}$ in $V_{\rm t}$.

\setcounter{table}{0}
\begin{table*}
\begin{minipage}{14cm}
\caption[]{Computed colour excesses for FGK supergiants.
    Negative E({\it B--V}) values are set to zero.}
\label{tab1}
\begin{tabular}{cclrrrrclrrr}
\hline
HD &$T_{\rm eff}$ &$\log g$ &$V_{\rm t}$ &[Fe/H] &E({\it B--V}) &HD/BD &$T_{\rm eff}$ &$\log g$ &$V_{\rm t}$ &[Fe/H] &E({\it B--V}) \\
\hline
000725&7053&  2.1 & 6.3 &--0.04&  0.271&171635&6151&  2.15& 5.2 &--0.04&  0.051 \\
001457&7636&  2.3 & 4.8 &--0.04&  0.394&172365&6196&  2.5 & 7.5 &--0.07&  0.180 \\
004362&5301&  1.6 & 4.4 &--0.15&  0.219&173638&7444&  2.4 & 4.7 &  0.11&  0.302 \\
007927&7341&  1.0 & 8.7 &--0.24&  0.425&174104&5657&  3.1 & 4.8 &--0.02&  0.045 \\
008890&6008&  2.2 & 4.35&  0.07&--0.007&179784&4956&  2.0 & 2.5 &  0.08&  0.394 \\
008906&6710&  2.2 & 4.8 &--0.07&  0.350&180028&6307&  1.9 & 4.0 &  0.10&  0.320 \\
009900&4529&  1.7:& 2.7 &  0.10&  0.073&182296&5072&  2.1 & 3.6 &  0.17&  0.321 \\
009973&6654&  2.0 & 5.7 &--0.05&  0.434&182835&6969&  1.6 & 4.9 &  0.00&  0.268 \\
010494&6672&  1.25& 7.5 &--0.20&  0.813&183864&5323&  1.8 & 3.5 &--0.02&  0.425 \\
011544&5126&  1.4 & 3.5 &  0.01&  0.174&185758&5367&  2.4 & 2.1 &--0.03&  0.037 \\
016901&5505&  1.7 & 4.3 &--0.03&  0.110&187203&5710&  2.2 & 5.1 &  0.05&  0.222 \\
017971&6822&  1.3 & 8.7 &--0.20&  0.644&187299&4566&  1.2 & 3.5 &  0.03&  0.257 \\
018391&5756&  1.2 &11.5 &  0.02&  0.991&187428&5892&  2.4:& 2.9 &  0.02&  0.177 \\
020902&6541&  2.0 & 4.8 &--0.01&  0.039&190113&4784&  1.9 & 3.5 &  0.05&  0.360 \\
025056&5752&  2.1 & 5.6 &  0.15&  0.437&190403&4894&  2.0 & 2.5 &  0.09&  0.133 \\
025291&7497&  2.65& 4.1 &  0.00&  0.241&191010&5253&  2.1:& 1.9 &  0.05&  0.171 \\
026630&5309&  1.8 & 3.7 &  0.02&  0.085&194093&6244&  1.7 & 6.1 &  0.05&  0.087 \\
032655&6755&  2.7 & 5.0 &--0.12&  0.059&195295&6572&  2.4 & 3.5 &  0.01&  0.001 \\
032655&6653&  2.5 & 5.0 &--0.13&  0.044&200102&5364&  1.6 & 3.3 &--0.13&  0.250 \\
036673&7500&  2.3 & 4.4 &  0.07&--0.046&200805&6865&  2.2 & 4.6 &--0.03&  0.455 \\
036891&5089&  1.7 & 3.3 &--0.06&  0.081&202314&5004&  2.1 & 3.2 &  0.12&  0.082 \\
039949&5239&  2.0:& 3.3 &--0.05&  0.233&202618&6541&  2.8 & 4.0 &--0.16&  0.053 \\
044391&4599&  1.6 & 3.4 &  0.03&  0.119&204022&5337&  1.5:& 3.9 &  0.01&  0.602 \\
045348&7557&  2.2 & 2.7 &--0.10&  0.016&204075&5262&  2.0 & 2.6 &--0.08&  0.186 \\
047731&4989&  2.0 & 3.2 &  0.02&  0.092&204867&5431&  1.6 & 4.15&--0.04&  0.006 \\
048329&4583&  1.2 & 3.7 &  0.16&  0.021&206859&4912&  1.2 & 2.5 &  0.04&  0.065 \\
052497&5090&  2.45& 3.6 &--0.02&  0.033&207489&6350&  2.85& 5.6 &  0.13&  0.119 \\
054605&6364&  1.5 &10.2 &--0.03&  0.085&208606&4702&  1.4 & 4.0 &  0.11&  0.323 \\
057146&5126&  1.9 & 3.6 &  0.17&--0.019&209750&5199&  1.4 & 3.55&  0.02&  0.022 \\
061227&7433&  2.5 & 5.5 &--0.16&  0.284&210848&6238&  3.0 & 3.2 &  0.08&  0.001 \\
074395&5247&  1.8 & 3.0 &--0.01&--0.028&216206&5003&  2.1 & 3.2 &  0.02&  0.158 \\
077912&4975&  2.0 & 2.4 &  0.01&  0.061&218600&7458&  2.4:& 4.8 &--0.07&  0.653 \\
084441&5281&  2.15& 2.15&--0.01&  0.006&219135&5430&  1.75& 3.6 &--0.01&  0.296 \\
092125&5336&  2.4 & 2.7 &  0.05&  0.020&220102&6832&  2.5 & 5.8 &--0.23&  0.262 \\
159181&5214&  2.2 & 3.4 &  0.04&  0.087&223047&4864&  1.7 & 3.4 &  0.07&--0.005 \\
164136&6483&  3.1 & 4.5 &--0.37&  0.018&224165&4804&  1.9 & 2.5 &  0.08&  0.064 \\
171237&6792&  2.6 & 4.4 &--0.09&  0.175&+60 2532&6268&1.8 &5.2  &--0.01&  0.597 \\
\hline
\end{tabular}
\end{minipage}
\end{table*}

\section{Colour excesses}
Values of $T_{\rm eff}$, $\log g$, $V_{\rm t}$, and [Fe/H] obtained in the described manner can be used to determine intrinsic colours for the target Cepheids and FGK supergiants in our sample. As noted earlier, the effective temperature $T_{\rm eff}$ of a FGK supergiant can be linked to intrinsic ({\it B--V})$_0$ colour using relationships such as that of \citet{gr92}. For this study, however, a recalibration was made using unreddened ({\it B--V})$_0$ colours from \citet{be96} for supergiants and
from \citet{fe95}, \citet{ta03} and \citet{lc07} for Cepheids. Any disadvantages arising from combining different sources of reddening for the Cepheids appear to be negligible, as discussed in the following section. For Cepheids in the sample, an instantaneous ``observed" {\it B--V} colour index was obtained from the extensive database of \citet{be07}, which contains multicolour photoelectric observations for all of our 164 program Cepheids. Published ephemerides were used to phase the data, and the light curves were subjected to Fourier analysis, with coefficients determined up to the third to tenth order, in order to match them (see Figs. 2 and 3).

\begin{figure}
\includegraphics[width=8cm]{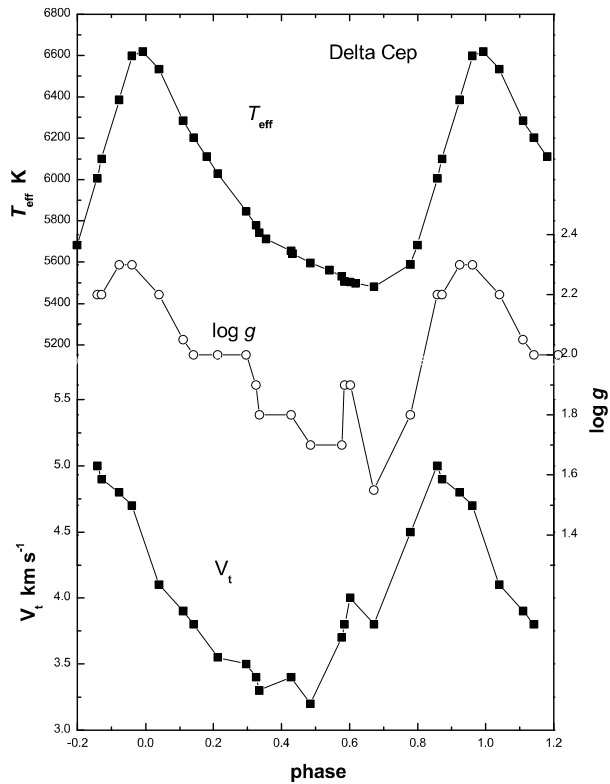}
\caption{The variation of $T_{\rm eff}$, $\log g$, and $V_{\rm t}$ for $\delta$ Cep.}
\label{fig1}
\end{figure}

\begin{figure}
\includegraphics[width=8cm]{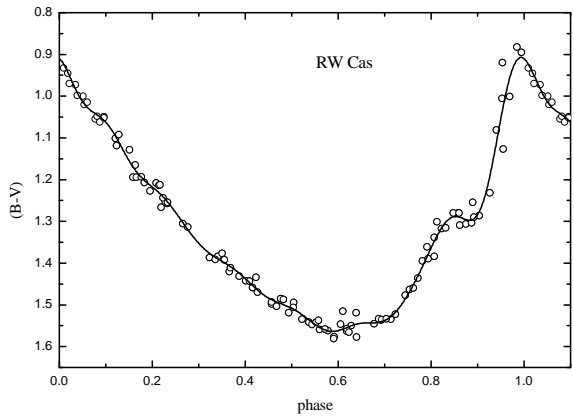}
\caption{{\it B--V} colour curve for RW Cas. The Fourier fit is shown as a line.}
\label{fig2}
\end{figure}

\begin{figure}
\includegraphics[width=8cm]{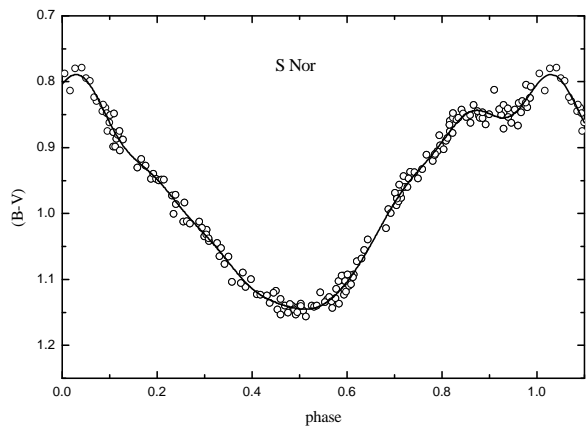}
\caption{As in Fig. 2 but for S Nor.}
\label{fig3}
\end{figure}

Published ephemerides were also used to derive pulsation phases for the times of the individual spectroscopic observations. In such fashion a pair of reddened and unreddened {\it B--V} values was generated for each epoch of spectroscopic observation for a Cepheid. Since the colour excess should not vary during a pulsation cycle, multivariate least squares can be used to link the observed intrinsic parameters for Cepheids and supergiants to intrinsic ({\it B--V})$_0$ colour.

Problems can arise when comparing photometric and spectroscopic data from different epochs, particularly for long period Cepheids which undergo very rapid changes in pulsation period relative to Cepheids of short period \citep[see][]{tu06a}. To minimize such problems, modern elements (quadratic, for example) from \citet{be07} were used to tie phases of spectroscopic observation to those for nearby (close) epochs of photometric observation. But it was not always possible to minimize the differences between epochs of photometric and spectroscopic observation for all Cepheids, with southern hemisphere objects remaining as a potential source of error.

In summary, the method takes known atmospheric parameters ($T_{\rm eff}$, $\log g$, $V_{\rm t}$, [Fe/H]) for the supergiants and Cepheids, and adopts a colour excess E({\it B--V}) from \citet{be96}, \citet{fe95}, \citet{ta03}, and \citet{lc07}. For objects common to those studies, the values were averaged. An observed value of {\it B--V} for the supergiant or Cepheid was obtained next, in the case of Cepheids using data from \citet{be07} phased in the same manner as the spectroscopic data, as noted above. The analysis therefore yields ({\it B--V})$_0$ for the supergiants, and comparable values for the Cepheids at the observed phases. A multivariate regression of ($T_{\rm eff}$, $\log g$, $V_{\rm t}$, [Fe/H]) versus ({\it B--V})$_0$ was then performed on the 693 individual observations to produce the following relationship:
\begin{center}
$(B-V)_{0}=(57.984 \pm4.485)-(10.3587\pm0.9797)(\log T_{\rm eff})^{2}+(1.67572\pm0.17631)(\log T_{\rm eff})^{3}-(3.356\pm0.461)\log g+(0.0321 \pm0.0024)V_{\rm t}+(0.2615 \pm0.0301)[{\rm Fe/H}]+
(0.8833 \pm 0.1229)(\log g)(\log T_{\rm eff}).$
\end{center}

With the above relationship it was possible to derive homogeneous colour excesses E({\it B--V}) = ({\it B--V}) -- ({\it B--V})$_0$ for 74 FGK supergiants and 164 classical Cepheids. The former are presented in Table 1, the latter are summarized in Table 2. The precision in the final estimates of ({\it B--V})$_0$ is estimated to be 0.04--0.05 mag (1 sigma, external precision) for spectra of R = 42\,000, S/N = 100--150. The internal precision is 0.0025 mag. The results could be improved further with the addition of higher resolution and larger S/N ratio spectra. We note that the error budget does not include possible uncertainties that arise from the individual properties of stars, such as rotation, chemical composition, binarity, etc.

\setcounter{table}{1}
\begin{table*}
\begin{minipage}{14cm}
\caption{Computed colour excesses for classical Cepheids.}
\label{tab2}
\begin{tabular}{lrcccccc}
\hline
Name &$P$ (days) &E({\it B--V}) &$\pm$ &No. Spectra &E({\it B--V})$_{Fer95}$ &E({\it B--V})$_{TSR03}$ &E({\it B--V})$_{LC07}$ \\
\hline
  U  Aql&   7.023958&   0.416 & ...   &  1&  0.401 &  0.371 &  0.364 \\
  SZ Aql&  17.140849&   0.588 & 0.041 & 11&  0.588 &  0.552 &  0.531 \\
  TT Aql&  13.754707&   0.480 & 0.036 &  8&  0.518 &  0.462 &  0.432 \\
  FF Aql&   4.470916&   0.224 & 0.017 & 14&  0.208 &  ...   &  0.191 \\
  FM Aql&   6.114230&   0.636 & 0.010 &  3&  0.676 &  0.617 &  0.595 \\
  FN Aql&   9.481603&   0.469 & 0.016 &  5&  0.517 &  0.490 &  0.488 \\
V496 Aql&   6.807055&   0.408 & 0.009 &  3&  0.453 &  ...   &  0.390 \\
V600 Aql&   7.238748&   0.830 & ...   &  1&  0.864 &  0.819 &  0.792 \\
V733 Aql&   6.178748&   0.106 & ...   &  1&  ...   &  ...   &  ...   \\
V1162 Aql&  5.376100&   0.163 & ...   &  1&  0.196 &  0.187 &  ...   \\
V1344 Aql&  7.478030&   0.544 & ...   &  2&  0.626 &  0.569 &  ...   \\
 Eta Aql&   7.176735&   0.096 & 0.015 & 14&  0.152 &  0.133 &  0.138 \\
V340 Ara&  20.809000&   0.540 & ...   &  1&  0.583 &  0.546 &  ...   \\
  RT Aur&   3.728190&   0.050 & 0.036 & 10&  0.063 &  0.049 &  0.089 \\
  SY Aur&  10.144698&   0.457 & ...   &  1&  0.411 &  0.453 &  ...   \\
  YZ Aur&  18.193212&   0.580 & 0.030 &  5&  0.583 &  0.601 &  0.575 \\
  AN Aur&  10.290560&   0.559 & 0.051 &  3&  0.577 &  0.600 &  ...   \\
  BK Aur&   8.002432&   0.396 & ...   &  1&  0.418 &  0.424 &  ...   \\
  CY Aur&  13.847650&   0.905 & ...   &  1&  0.837 &  0.767 &  ...   \\
  ER Aur&  15.690730&   0.618 & ...   &  1&  0.491 &  0.494 &  ...   \\
V335 Aur&   3.413250&   0.701 & ...   &  1&  0.658 &  0.626 &  ...   \\
  RY CMa&   4.678250&   0.232 & ...   &  1&  0.253 &  0.223 &  0.258 \\
  TW CMa&   6.995070&   0.311 & ...   &  1&  0.398 &  0.391 &  0.280 \\
  VZ CMa&   3.126230&   0.537 & ...   &  1&  0.474 &  ...   &  ...   \\
  RX Cam&   7.912024&   0.539 & 0.034 &  9&  0.581 &  0.536 &  0.553 \\
  TV Cam&   5.294970&   0.482 & ...   &  1&  0.596 &  0.612 &  ...   \\
  AB Cam&   5.787640&   0.681 & ...   &  1&  0.662 &  0.673 &  ...   \\
  AD Cam&  11.260991&   0.939 & ...   &  1&  0.929 &  0.871 &  ...   \\
  RW Cas&  14.791548&   0.415 & ...   &  2&  0.468 &  0.409 &  0.383 \\
  RY Cas&  12.138880&   0.580 & ...   &  1&  0.676 &  0.613 &  ...   \\
  SU Cas&   1.949322&   0.296 & 0.026 & 13&  0.288 &  ...   &  0.282 \\
  SW Cas&   5.440950&   0.517 & ...   &  1&  0.505 &  0.449 &  0.484 \\
  SY Cas&   4.071098&   0.449 & ...   &  1&  0.478 &  0.430 &  ...   \\
  SZ Cas&  13.637747&   0.949 & ...   &  1&  0.759 &  ..   .&  0.809 \\
  XY Cas&   4.501697&   0.494 & ...   &  1&  0.550 &  0.480 &  ...   \\
  BD Cas&   3.650900&   1.006 & ...   &  1&  ...   &  ...   &  ...   \\
 CEa Cas&   5.141058&   0.503 & ...   &  1&  0.591 &  0.562 &  ...   \\
 CEb Cas&   4.479301&   0.479 & ...   &  1&  0.576 &  0.548 &  ...   \\
  CF Cas&   4.875220&   0.527 & 0.025 &  5&  0.591 &  0.531 &  0.544 \\
  CH Cas&  15.086190&   0.907 & ...   &  1&  1.002 &  0.955 &  ...   \\
  CY Cas&  14.376860&   0.891 & ...   &  1&  1.013 &  0.947 &  ...   \\
  DD Cas&   9.812027&   0.439 & ...   &  1&  0.517 &  0.493 &  0.446 \\
  DL Cas&   8.000669&   0.487 & 0.024 & 14&  0.518 &  0.479 &  0.499 \\
  FM Cas&   5.809284&   0.324 & ...   &  1&  0.350 &  0.290 &  0.309 \\
V379 Cas&   4.305750&   0.600 & ...   &  1&  ...   &  ...   &  ...   \\
V636 Cas&   8.377000&   0.553 & 0.012 &  8&  0.631 &  ...   &  ...   \\
  V  Cen&   5.493861&   0.167 & ...   &  1&  0.282 &  0.264 &  0.311 \\
  CP Cep&  17.859000&   0.659 & ...   &  1&  0.724 &  0.702 &  ...   \\
  CR Cep&   6.232964&   0.721 & ...   &  1&  0.749 &  0.697 &  0.698 \\
  IR Cep&   2.114124&   0.368 & ...   &  1&  0.434 &  ...   &  ...   \\
V351 Cep&   2.805910&   0.436 & ...   &  1&  ...   &  ...   &  ...   \\
Delta Cep&  5.366270&   0.045 & 0.018 & 18&  0.080 &  0.068 &  0.087 \\
  BG Cru&   3.342720&   0.087 & ...   &  1&  0.111 &  ...   &  ...   \\
  X  Cyg&  16.386332&   0.239 & 0.029 & 26&  0.267 &  0.261 &  0.208 \\
  SU Cyg&   3.845492&   0.074 & 0.019 & 12&  0.133 &  0.088 &  0.091 \\
  SZ Cyg&  15.109642&   0.576 & ...   &  1&  0.632 &  0.587 &  0.562 \\
  TX Cyg&  14.708157&   1.112 & ...   &  1&  1.195 &  1.111 &  1.179 \\
  VX Cyg&  20.133407&   0.830 & ...   &  1&  0.889 &  0.830 &  ...   \\
  VY Cyg&   7.856982&   0.639 & ...   &  1&  0.634 &  0.615 &  0.597 \\
  VZ Cyg&   4.864453&   0.246 & ...   &  1&  0.310 &  0.274 &  0.270 \\
  BZ Cyg&  10.141932&   0.872 & ...   &  1&  0.885 &  0.839 &  ...   \\
  CD Cyg&  17.073967&   0.447 & 0.040 & 16&  0.545 &  0.486 &  0.513 \\
  DT Cyg&   2.499082&   0.028 & 0.009 & 14&  0.067 &  ...   &  0.048 \\
\hline
\end{tabular}
\end{minipage}
\end{table*}

\setcounter{table}{1}
\begin{table*}
\begin{minipage}{14cm}
\caption{Continued.}
\label{tab2}
\begin{tabular}{lrcccccc}
\hline
Name &$P$ (days) &E({\it B--V}) &$\pm$ &No. Spectra &E({\it B--V})$_{Fer95}$ &E({\it B--V})$_{TSR03}$ &E({\it B--V})$_{LC07}$ \\
\hline
  MW Cyg&   5.954586&   0.692 & ...   &  1&  0.693 &  0.615 &  0.645 \\
V386 Cyg&   5.257606&   0.891 & ...   &  1&  0.965 &  0.884 &  0.844 \\
V402 Cyg&   4.364836&   0.366 & ...   &  1&  0.484 &  0.397 &  0.391 \\
V532 Cyg&   3.283612&   0.568 & ...   &  1&  0.552 &  ...   &  0.453 \\
V924 Cyg&   5.571472&   0.233 & ...   &  1&  0.258 &  ...   &  0.256 \\
V1154 Cyg&  4.925460&   0.234 & ...   &  1&  0.335 &  ...   &  ...   \\
V1334 Cyg&  3.333020&   0.025 & 0.009 & 11&--0.055 &  ...   &  ...   \\
V1726 Cyg&  4.237060&   0.318 & ...   &  1&  0.312 &  ...   &  0.361 \\
  TX Del&   6.165907&   0.222 & ...   &  1&        &  ...   &  ...   \\
Beta Dor&   9.842425&   0.000 & ...   &  1&  0.080 &  0.069 &  0.041 \\
  W  Gem&   7.913779&   0.273 & 0.040 & 11&  0.281 &  0.266 &  0.252 \\
  AA Gem&  11.302328&   0.231 & ...   &  1&  0.367 &  0.380 &  0.264 \\
  AD Gem&   3.787980&   0.164 & ...   &  1&  0.219 &  0.159 &  0.144 \\
  DX Gem&   3.137486&   0.507 & ...   &  1&  0.462 &  ...   &  0.430 \\
Zeta Gem&  10.150135&   0.031 & 0.041 & 11&  0.044 &  0.033 &  0.009 \\
  V  Lac&   4.983468&   0.417 & ...   &  1&  0.312 &  0.315 &  0.338 \\
  X  Lac&   5.444990&   0.350 & ...   &  1&  0.330 &  0.339 &  0.336 \\
  Y  Lac&   4.323776&   0.136 & 0.011 &  9&  0.226 &  0.202 &  0.195 \\
  Z  Lac&  10.885613&   0.415 & 0.043 &  9&  0.375 &  0.378 &  0.368 \\
  RR Lac&   6.416243&   0.363 & ...   &  1&  0.284 &  0.296 &  0.353 \\
  BG Lac&   5.331932&   0.272 & 0.020 &  3&  0.358 &  0.316 &  0.287 \\
V473 Lyr&   1.490780&   0.091 & ...   &  1&  0.049 &  ...   &  ...   \\
  T  Mon&  27.024649&   0.179 & 0.029 & 19&  0.221 &  0.195 &  0.188 \\
  SV Mon&  15.232780&   0.188 & 0.034 & 11&  0.281 &  0.250 &  0.214 \\
  TW Mon&   7.096900&   0.682 & ...   &  1&  0.696 &  ...   &  ...   \\
  TX Mon&   8.701731&   0.457 & ...   &  1&  0.503 &  0.492 &  0.465 \\
  TZ Mon&   7.428014&   0.448 & ...   &  1&  0.462 &  0.431 &  ...   \\
  UY Mon&   2.397970&   0.163 & ...   &  1&  0.108 &  ...   &  ...   \\
  WW Mon&   4.662310&   0.688 & ...   &  1&  0.602 &  0.561 &  ...   \\
  XX Mon&   5.456473&   0.610 & ...   &  1&  0.623 &  0.586 &  ...   \\
  AA Mon&   3.938164&   0.724 & ...   &  1&  0.814 &  0.791 &  ...   \\
  AC Mon&   8.014250&   0.579 & ...   &  1&  0.750 &  0.702 &  0.776 \\
  EE Mon&   4.808960&   0.509 & ...   &  1&  0.539 &  0.507 &  ...   \\
  CU Mon&   4.707873&   0.859 & ...   &  1&  0.792 &  0.750 &  ...   \\
  CV Mon&   5.378898&   0.681 & ...   &  1&  0.488 &  0.464 &  ...   \\
  EK Mon&   3.957941&   0.518 & ...   &  1&  0.582 &  0.551 &  ...   \\
  FG Mon&   4.496590&   0.625 & ...   &  1&  0.684 &  0.650 &  ...   \\
  FI Mon&   3.287822&   0.607 & ...   &  1&  0.539 &  0.513 &  ...   \\
V465 Mon&   2.713176&   0.142 & ...   &  1&  0.263 &  0.255 &  ...   \\
V504 Mon&   2.774050&   0.588 & ...   &  1&  0.565 &  ...   &  ...   \\
V508 Mon&   4.133608&   0.221 & ...   &  1&  0.330 &  0.320 &  ...   \\
V526 Mon&   2.674985&   0.218 & ...   &  1&  0.093 &  ...   &  ...   \\
  S  Nor&   9.754244&   0.268 & ...   &  1&  0.194 &  0.178 &  0.188 \\
  Y  Oph&  17.126908&   0.683 & 0.010 & 14&  0.645 &  ...   &  0.668 \\
  BF Oph&   4.067510&   0.270 & ...   &  1&  0.278 &  0.247 &  0.223 \\
  RS Ori&   7.566881&   0.410 & ...   &  1&  0.353 &  0.335 &  0.350 \\
  GQ Ori&   8.616068&   0.272 & ...   &  1&  0.238 &  0.228 &  0.268 \\
  SV Per&  11.129318&   0.304 & ...   &  1&  0.345 &  0.366 &  0.431 \\
  UX Per&   4.565815&   0.522 & ...   &  1&  0.492 &  0.512 &  ...   \\
  VX Per&  10.889040&   0.486 & 0.027 &  8&  0.508 &  0.496 &  0.477 \\
  AS Per&   4.972516&   0.685 & ...   &  1&  0.728 &  0.644 &  0.685 \\
  AW Per&   6.463589&   0.515 & 0.015 &  4&  0.510 &  0.487 &  0.476 \\
  BM Per&  22.951900&   0.975 & 0.016 &  4&  0.978 &  0.870 &  ...   \\
  HQ Per&   8.637930&   0.543 & ...   &  1&  0.571 &  ...   &  ...   \\
  MM Per&   4.118415&   0.480 & ...   &  1&  0.515 &  0.490 &  ...   \\
V440 Per&   7.570000&   0.283 & 0.036 & 10&  0.274 &  ...   &  ...   \\
  X  Pup&  25.961000&   0.429 & 0.033 &  7&  0.421 &  0.409 &  0.399 \\
  RS Pup&  41.387600&   0.515 & ...   &  1&  0.480 &  0.453 &  0.480 \\
  VZ Pup&  23.171000&   0.672 & ...   &  1&  0.461 &  0.452 &  0.424 \\
  AD Pup&  13.594000&   0.222 & ...   &  1&  0.386 &  0.343 &  ...   \\
  AQ Pup&  30.104000&   0.453 & ...   &  1&  0.565 &  0.531 &  0.504 \\
  BN Pup&  13.673100&   0.425 & ...   &  1&  0.449 &  0.417 &  0.415 \\
  HW Pup&  13.454000&   0.706 & ...   &  1&  0.662 &  0.688 &  ...   \\
\hline
\end{tabular}
\end{minipage}
\end{table*}

\setcounter{table}{1}
\begin{table*}
\begin{minipage}{14cm}
\caption{Continued.}
\label{tab2}
\begin{tabular}{lrcccccc}
\hline
Name &$P$ (days) &E({\it B--V}) &$\pm$ &No. Spectra &E({\it B--V})$_{Fer95}$ &E({\it B--V})$_{TSR03}$ &E({\it B--V})$_{LC07}$ \\
\hline
  MY Pup&   5.695309&   0.105 & ...   &  1&  0.112 &  ...   &  ...   \\
  RV Sco&   6.061306&   0.329 & ...   &  1&  0.365 &  0.338 &  0.381 \\
  RY Sco&  20.320144&   0.774 & ...   &  1&  0.696 &  0.714 &  0.760 \\
  KQ Sco&  28.689600&   0.893 & ...   &  1&  0.906 &  0.839 &  0.911 \\
V500 Sco&   9.316839&   0.668:& 0.158 &  9&  0.603 &  0.568 &  0.621 \\
  Z  Sct&  12.901325&   0.560 & ...   &  1&  0.569 &  0.491 &  0.534 \\
  S  Sge&   8.382086&   0.116 & 0.016 & 15&  0.128 &  0.112 &  0.099 \\
  SS Sct&   3.671253&   0.391 & ...   &  1&  0.362 &  0.317 &  0.333 \\
  UZ Sct&  14.744200&   0.986 & ...   &  1&  1.020 &  0.973 &  ...   \\
  EV Sct&   3.090990&   0.737 & ...   &  1&  0.663 &  ...   &  0.650 \\
  U  Sgr&   6.745229&   0.398 & 0.022 &  9&  0.434 &  0.403 &  0.421 \\
  W  Sgr&   7.594904&   0.079 & 0.017 &  8&  0.116 &  0.112 &  0.108 \\
  X  Sgr&   7.012877&   0.219 & ...   &  2&  0.201 &  0.201 &  0.230 \\
  Y  Sgr&   5.773380&   0.182 & 0.021 & 12&  0.216 &  0.188 &  0.195 \\
  VY Sgr&  13.557200&   1.136 & ...   &  1&  0.961 &  1.143 &  ...   \\
  WZ Sgr&  21.849789&   0.458 & 0.058 & 12&  0.486 &  0.428 &  0.467 \\
  XX Sgr&   6.424140&   0.575 & ...   &  1&  0.524 &  ...   &  0.549 \\
  YZ Sgr&   9.553687&   0.282 & 0.015 &  8&  0.307 &  0.285 &  0.298 \\
  AP Sgr&   5.057916&   0.186 & ...   &  1&  0.196 &  0.174 &  0.195 \\
  AV Sgr&  15.415000&   1.117 & ...   &  1&  1.317 &  1.233 &  ...   \\
  BB Sgr&   6.637102&   0.287 & ...   &  1&  0.303 &  0.276 &  0.315 \\
V350 Sgr&   5.154178&   0.280 & ...   &  1&  0.328 &  0.295 &  0.330 \\
  ST Tau&   4.034299&   0.306 & 0.076 &  3&  0.349 &  0.339 &  0.393 \\
  SZ Tau&   3.148380&   0.272 & 0.022 & 18&  0.326 &  ...   &  0.308 \\
  AE Tau&   3.896450&   0.576 & ...   &  1&  0.604 &  ...   &  ...   \\
  EF Tau&   3.448150&   0.360 & ...   &  1&  ...   &  ...   &  ...   \\
  EU Tau&   2.102480&   0.230 & ...   &  1&  0.184 &  ...   &  ...   \\
 Alp UMi&   3.969600& --0.010 & ...   &  1&  0.017 &  ...   &  0.009 \\
  T  Vel&   4.639819&   0.272 & ...   &  1&  0.300 &  0.271 &  0.286 \\
  RY Vel&  28.135700&   0.602 & ...   &  1&  0.573 &  0.554 &  0.540 \\
  RZ Vel&  20.398240&   0.293 & ...   &  1&  0.320 &  0.293 &  0.283 \\
  SW Vel&  23.441000&   0.409 & 0.035 &  3&  0.360 &  0.337 &  0.335 \\
  SX Vel&   9.549930&   0.272 & ...   &  1&  0.252 &  0.250 &  0.270 \\
  S  Vul&  68.464000&   0.940 & 0.051 &  6&  0.782 &  0.737 &  0.674 \\
  T  Vul&   4.435462&   0.068 & 0.015 & 20&  0.098 &  0.067 &  0.054 \\
  U  Vul&   7.990629&   0.663 & 0.018 &  7&  0.636 &  0.593 &  0.640 \\
  X  Vul&   6.319543&   0.798 & 0.022 &  6&  0.824 &  0.790 &  0.702 \\
  SV Vul&  44.994772&   0.510 & 0.020 & 23&  0.504 &  0.518 &  0.412 \\
\hline
\end{tabular}
\end{minipage}
\\
\end{table*}

\begin{figure}
\includegraphics[width=8cm]{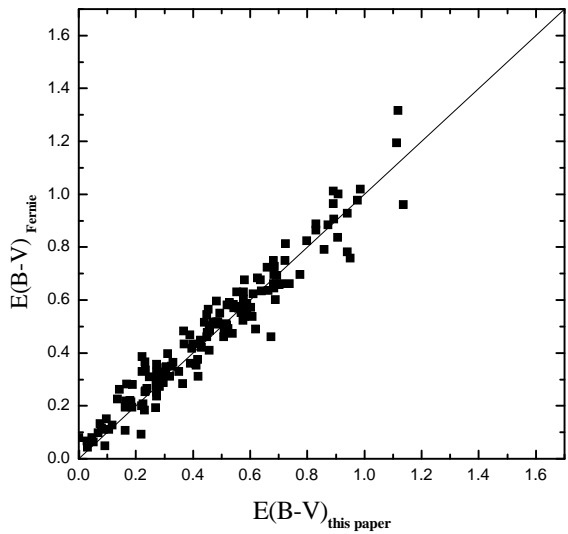}
\caption{Comparison of our colour excesses with those of \citet{fe95}.}
\label{fig4}
\end{figure}

\begin{figure}
\includegraphics[width=8cm]{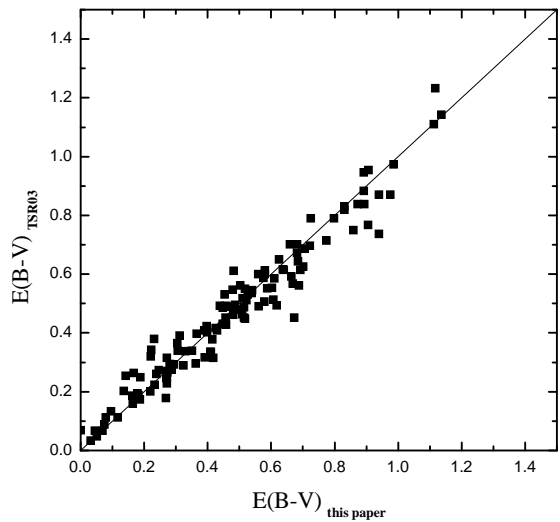}
\caption{Comparison of the present colour excesses with those of \citet{ta03}.}
\label{fig5}
\end{figure}

\begin{figure}
\includegraphics[width=8cm]{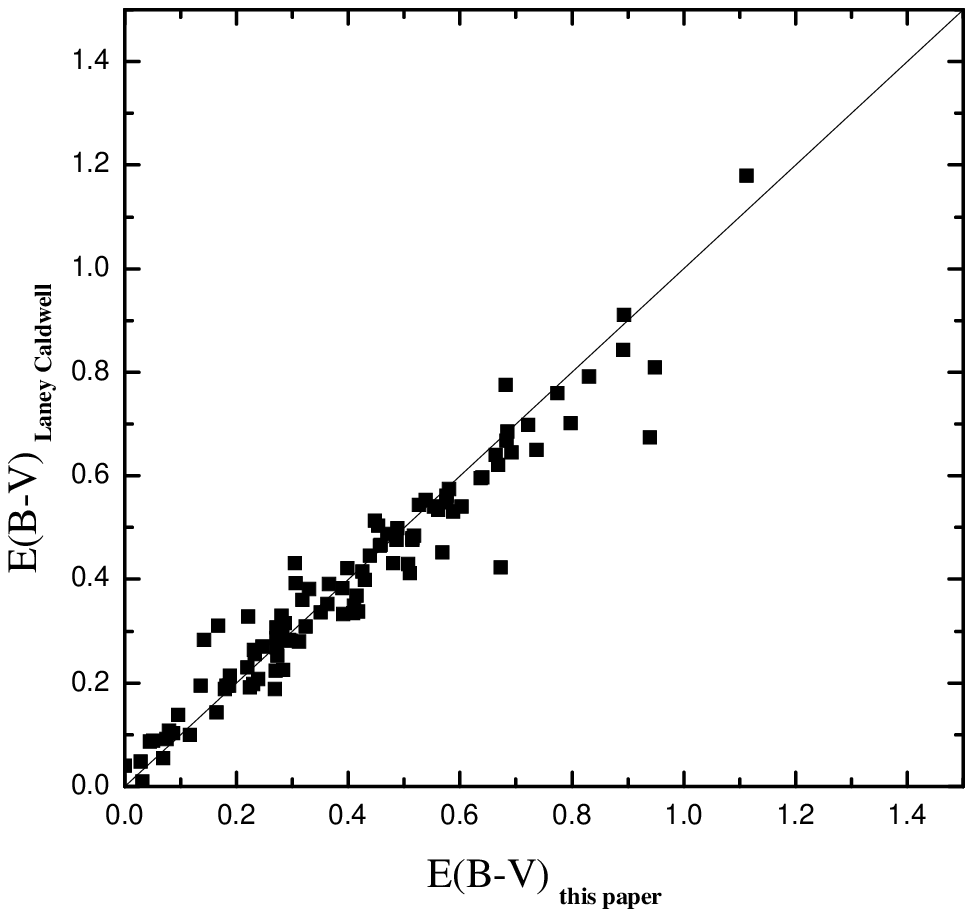}
\caption{Comparison of the present colour excesses with those of \citet{lc07}.}
\label{fig6}
\end{figure}

\begin{figure}
\includegraphics[width=8cm]{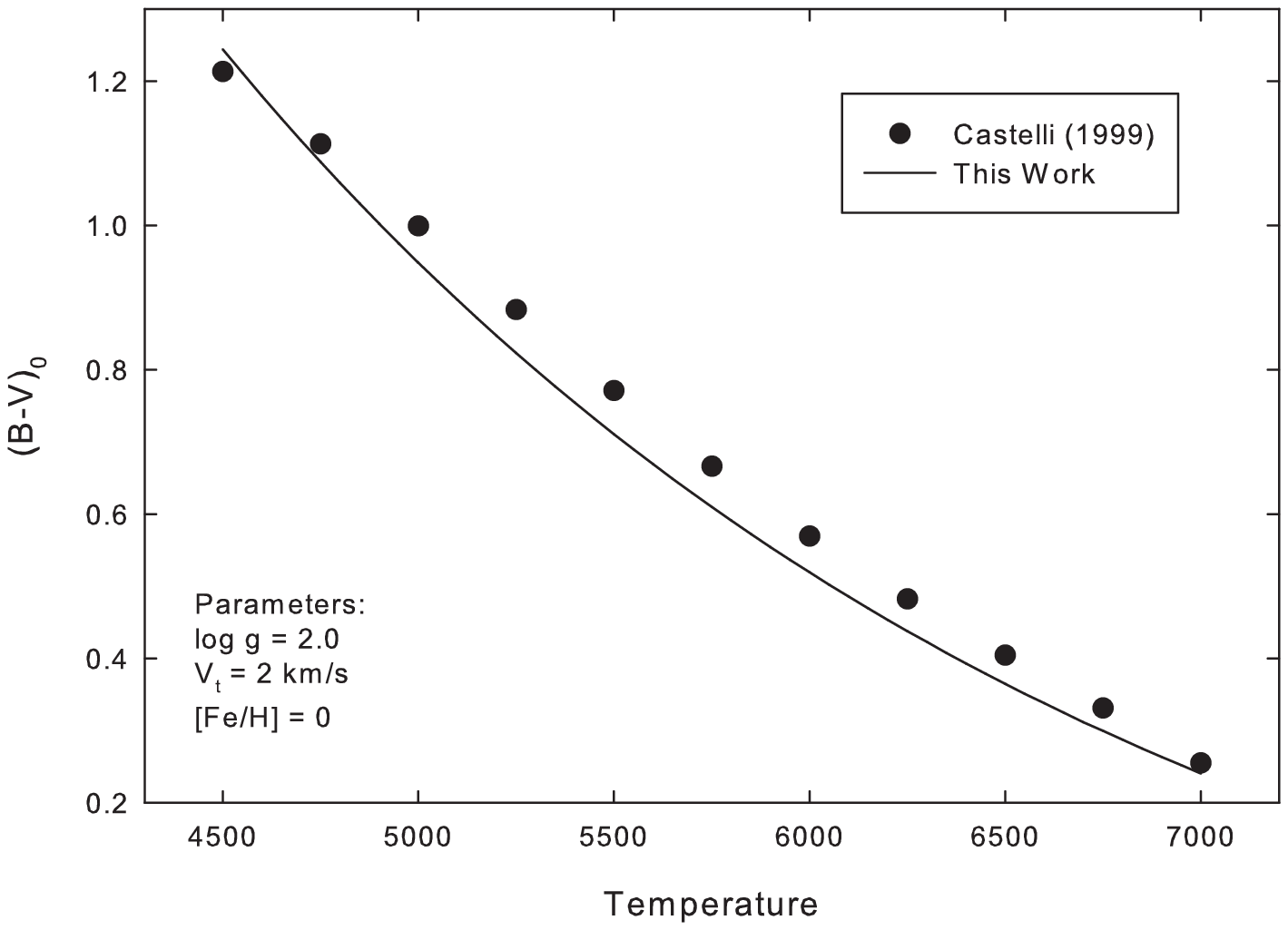}
\caption{Comparison of the present colours with those of \citet{ca99}.}
\label{fig7}
\end{figure}

\section{Results}
Tables 1 and 2 list the results for FGK supergiants and classical Cepheids, respectively. For Cepheids the variable star designation and pulsation period $P$ are listed in columns (1) and (2) respectively, the mean colour excess determined here is listed in column (3), the standard error of the mean is listed in column (4), and the number of determinations used to calculate the mean is listed in column (5). The mean value of E({\it B--V}) was obtained by averaging the values of E({\it B--V}) over the pulsation cycle. Individual reddenings estimated by \citet{fe95}, \citet{ta03}, and \citet{lc07} are also listed for convenience in Table 2, in columns (5), (6), and (7) respectively. A comparison of the reddenings of \citet{fe95} with the present results is shown in Fig. 4, and comparisons with the reddenings of \citet{ta03} and \citet{lc07} are shown in Figs. 5 and 6.

A statistical comparison of the present Cepheid reddenings was made with an overlapping set of space reddenings for 27 stars in common. The space reddenings are those summarized by \citet{lc07}, but with additional data used in a previous analysis of Cepheid reddenings \citep{tu01} that includes new estimates for previously-unstudied objects \citep{tu08}. Four of the objects in the sample display large differences in reddening between the two data sets, in excess of $\Delta$E({\it B--V}) = 0.1, namely V Cen, S Nor, VY Sgr, and WZ Sgr. All are southern hemisphere Cepheids, noted to be a problem in the previous section, and the first three have only single spectroscopic observations available. Multiple observations ($n=12$) exist for WZ Sgr, but its space reddening has been noted previously to be a problem \citep{lc07}. For the 23 remaining Cepheids the difference in colour excesses $\Delta$E({\it B--V})(space reddening -- spectroscopic reddening) is $+0.005\pm0.039$ s.d., and $+0.000\pm0.035$ s.d. for the 16 Cepheids with multiple spectroscopic observations. It appears that the present compilation of spectroscopic reddenings for Cepheids is tied closely to the system of available space reddenings. In addition, since the average uncertainty for the space reddenings of individual Cepheids in the sample is $\pm0.024$, the precision of the spectroscopic reddenings based upon multiple spectra must be $\pm0.025$, while the accuracy matches that of the space reddening system.

The reddening system of the present study also appears to be closely matched to older reddening systems for Cepheids, as indicated by the trends of Figs. 4--6. The scatter in the plots of Figs. 4 and 5 is suggestive of actual differences that may exist, which may have several explanations. Period-colour relations, for example, do not always account for the intrinsic dispersion of Cepheids within the instability strip. Small amplitude Cepheids tend to lie on the hot edge of the strip, for example \citep{pl78,tu06a}, so their reddenings can be underestimated by period-colour relations. Large amplitude Cepheids lie near the centre of the strip \citep[e.g.,][]{tu06a}, but may not always be representative of the calibrators used. Although similar scatter is seen in the Fig. 6, where the present results are compared with the space reddening system of \citet{lc07}, the discrepant points in most cases represent Cepheids for which only single spectra are available. Here the problem is not of differences in reddening systems, but problems in the measurements for individual Cepheids.

All Cepheids undergo period changes \citep{tu06a}, some undergo random fluctuations in period \citep{be04,ba07}, and others display light travel time effects from binarity and other curious changes to the times of light maximum \citep{tu07} that affect the accurate determination of light curve phasing, all of which are important factors for reddenings derived from spectroscopic and photometric observations. As noted previously, the problem is particularly acute for the present study when only one spectrum has been obtained \citep[see][]{tu06b}, which is the case for many of the Cepheids in the present sample. A number of objects in Table 2 are also suspected to be Type II Cepheids, but their [Fe/H] values are nearly solar according to the spectroscopic analyses.

Another comparison that can be made is with respect to colours derived from model atmospheres. Such a comparison is shown in Fig. 7, where the models (solid dots) are \citet{ku92} models with the colours
given by \citet{ca99}. The models are for $\log g$ = 2.0, $V_{\rm t}$ = 2.0 km s$^{-1}$, and [Fe/H] = 0.0. The same parameters were used to generate intrinsic colours from the calibration determined here.  As can be seen, the comparison is rather good. Over the larger part of the range the spectroscopic
temperatures used here are 50--100 K lower than the temperatures that would be determined using the photometric calibration. Another way of stating this is that, at a fixed temperature, the theoretical photometric calibration would yield a {\it B--V} colour excess about 0.05 mag. smaller than found here.
The difference between spectroscopic and photometric temperatures found here for high luminosity stars is in the same sense and magnitude as that found for spectroscopic versus photometric temperatures of dwarfs and giants \citep{lh06,lh07}.

It should be emphasized that high-resolution spectroscopic studies of Cepheids provide an independent source of reddenings that can be used to refine the existing reddening scale for Galactic Cepheids and to enhance the precision of the Cepheid P--L relation. The colour excesses derived for Cepheids in our sample for which many spectra are available are quite likely more accurate than many previous estimates, and provide a good test of published reddening scales. Additional spectroscopic observations of Cepheids in our sample can only strengthen the present results, since they would eliminate the undersampling problem inherent to program obejects with $n=1$. Correcting that deficiency should improve the precision of the overall sample.

\section{Summary}

The present study presents newly-derived parameters, namely effective temperature ($T_{\rm eff}$), surface gravity ($\log g$), microturbulent velocity ($V_{\rm t}$, in km s$^{-1}$), and iron abundance ([Fe/H]), for 74 non-variable FGK supergiants, established from model atmosphere analyses of high resolution spectra for the stars. Colour excesses have been computed for the stars on the basis of a new formulation of the relationship linking such parameters to intrinsic colour, ({\it B--V})$_0$. The formulation has been extended to the 74 FGK supergiants and to a sample of 164 classical Cepheid variables to derive new estimates for their colour excesses E({\it B--V}), presented in Tables 1 and 2.
The reddening estimates are demonstrated to be of extremely high internal precision, and to agree well with the most accurate estimates from the literature. The Cepheid reddenings, in particular, appear to be closely tied to the system of space reddenings that is presently available. Given the large distances of supergiants, the method opens the possibility for large-scale extinction mapping of the Galaxy, with a sensitivity of 0.08--0.2 magnitude.

\subsection*{ACKNOWLEDGEMENTS}
This work is based on spectra collected with the 1.93-m telescope of the OHP (France) and the ESO Telescopes at the Paranal Observatory under programme ID266.D-5655. We thank the referee for useful suggestions that helped to improve the presentation.

\end{document}